\documentclass[aps,prd,floatfix,nofootinbib,showpacs,superscriptaddress,twocolumn]{revtex4-2}
\usepackage{color}
\usepackage{graphicx}
\usepackage{dcolumn}
\usepackage{bm}
\usepackage[utf8]{inputenc}
\usepackage{gensymb}
\usepackage{latexsym}
\usepackage{latexsym}
\usepackage{amssymb}
\usepackage{amsmath}
\usepackage{amsfonts}
\usepackage[mathscr,scaled=1.15]{urwchancal}
\DeclareFontFamily{OT1}{pzc}{}
\DeclareFontShape{OT1}{pzc}{m}{it}%
{<-> s * [1.15] pzcmi7t}{}
\DeclareMathAlphabet{\mathpzc}{OT1}{pzc}{m}{it}

\begin{document}


\title{Pion and Kaon box contribution to $a_{\mu}^{\text{HLbL}}$}

\author{Ángel Miramontes}
 \email{angel-aml@hotmail.com}
\affiliation{%
Instituto de F\'isica y Matem\'aticas, Universidad Michoacana de San Nicol\'as de Hidalgo, Morelia, Michoac\'an 58040, Mexico 
}%

\author{Adnan Bashir}
 \email{adnan.bashir@umich.mx}
\affiliation{%
Instituto de F\'isica y Matem\'aticas, Universidad Michoacana de San Nicol\'as de Hidalgo, Morelia, Michoac\'an 58040, Mexico 
}%

\author{Khépani Raya}
 \email{khepani@ugr.es}
\affiliation{%
 Departamento de F\'isica Te\'orica y del Cosmos, Universidad de Granada, E-18071, Granada, Spain
}%

\author{Pablo Roig}
 \email{proig@fis.cinvestav.mx }
\affiliation{%
Departamento de F\'isica, Centro de Investigaci\'on y de Estudios Avanzados del IPN,\\ Apdo. Postal 14-740,07000 Ciudad de M\'exico, Mexico
}%

\date{\today}

\begin{abstract}
We present an evaluation of the $\pi^\pm$ and $K^\pm$ box contributions to the hadronic light-by-light piece of the muon's anomalous magnetic moment, $a_\mu$. The calculation of the corresponding electromagnetic form factors (EFFs) is performed within a Dyson-Schwinger equations (DSE) approach to quantum chromodynamics. These form factors are calculated in the so-called rainbow-ladder (RL) truncation, following two different evaluation methods and, subsequently, in a  further improved approximation scheme which incorporates meson cloud effects. The results are mutually consistent, indicating that in the domain of relevance for $a_\mu$ the obtained EFFs are practically equivalent.  Our analysis yields the combined estimates of $a_\mu^{\pi^\pm-box}=-(15.6\pm 0.2)\times 10^{-11}$ and $a_\mu^{K^\pm-\text{box}}=-(0.48\pm 0.02)\times 10^{-11}$, in full agreement with results previously obtained within the DSE formalism and other contemporary estimates.
\end{abstract}

\maketitle


\section{Introduction}
There has been a renewed interest in the anomalous magnetic moment of the muon, $a_\mu$, after the first measurement from the new muon g-2 experiment at FNAL \cite{Muong-2:2021ojo}
\begin{equation}
   a_\mu^{\mathrm{FNAL}}=116592040(54)\times10^{-11}\,.
\end{equation}
 Combining it with the final average from the muon g-2 measurements at BNL \cite{Muong-2:2006rrc} yields
\begin{equation}
 a_\mu^{\mathrm{Exp}}=116592061(41)\times10^{-11}
\end{equation}
as the corresponding world average,  which has  a remarkable precision of $0.35$ parts per million.
On the other hand, the outcome of the Muon g-2 Theory Initiative for the Standard Model prediction of this quantity \cite{Aoyama:2020ynm}~\footnote{This result is based on Refs. \cite{Davier:2017zfy, Keshavarzi:2018mgv, Colangelo:2018mtw, Hoferichter:2019mqg, Davier:2019can, Keshavarzi:2019abf, Kurz:2014wya, FermilabLattice:2017wgj, Budapest-Marseille-Wuppertal:2017okr, RBC:2018dos,Giusti:2019xct,Shintani:2019wai,FermilabLattice:2019ugu,Gerardin:2019rua,Aubin:2019usy,Giusti:2019hkz,Melnikov:2003xd,Masjuan:2017tvw,Colangelo:2017fiz,Hoferichter:2018kwz,Gerardin:2019vio,Bijnens:2019ghy,Colangelo:2019uex,Pauk:2014rta,Danilkin:2016hnh,Jegerlehner:2017gek,Knecht:2018sci,Eichmann:2019bqf,Roig:2019reh,Colangelo:2014qya,Blum:2019ugy,Aoyama:2012wk,Aoyama:2019ryr,Czarnecki:2002nt,Gnendiger:2013pva}. Later developments, after the cut for inclusion in the white paper \cite{Aoyama:2020ynm}, include Refs. \cite{Giusti:2020efo,Knecht:2020xyr,Masjuan:2020jsf,Ludtke:2020moa,Miranda:2020wdg,Hoid:2020xjs,Ananthanarayan:2020vum,Aubin:2020scy,Bijnens:2020xnl,Qin:2020udp,Bijnens:2021jqo,Zanke:2021wiq,Chao:2021tvp,Danilkin:2021icn,Colangelo:2021nkr,Yi:2021ccc,Leutgeb:2021mpu,James:2021sor, Colangelo:2021moe,Hoferichter:2021wyj}.},
\begin{equation}
    a_\mu^{\mathrm{SM}}=116591810(43)\times10^{-11}\,,
\end{equation}
has a comparable accuracy and deviates by $4.2\sigma$ from $a_\mu^{\mathrm{exp}}$. This  difference hints at the tantalizing prospect of new physics  being at work.

The BMW lattice QCD result~\cite{Borsanyi:2020mff} for the dominant component of the SM uncertainty, i.e., the hadronic vacuum polarization contribution (HVP) was not  taken into account for the $a_\mu^{\mathrm{SM}}$ value in the white paper~\cite{Aoyama:2020ynm}. If this value 
 were used, the incompatibility between the SM prediction and the world average  is reduced to merely $1.6\sigma$ level~\footnote{If the BMW result is adopted for $a_\mu^{\mathrm{HVP}}$, then there must anyway be new physics somewhere else, according to the constraints set by the electroweak precision observables~\cite{Crivellin:2020zul,Keshavarzi:2020bfy,deRafael:2020uif,Malaescu:2020zuc,Colangelo:2020lcg}.}. This result needs to be confirmed or refuted by other lattice analyses achieving a similar accuracy,  thus constituting a very active area of research.

The uncertainty in $a_\mu^{\mathrm{Exp}}$ is expected to shrink further as more data is analyzed, demanding a commensurate improvement in $a_\mu^{\mathrm{SM}}$. Its different contributions - particularly the hadronic ones (HVP and hadronic light-by-light, HLbL), which dominate the uncertainty- are continuously refined  to achieve this objective.

\begin{figure*}[t!]
\centerline{%
\includegraphics[width=13cm]{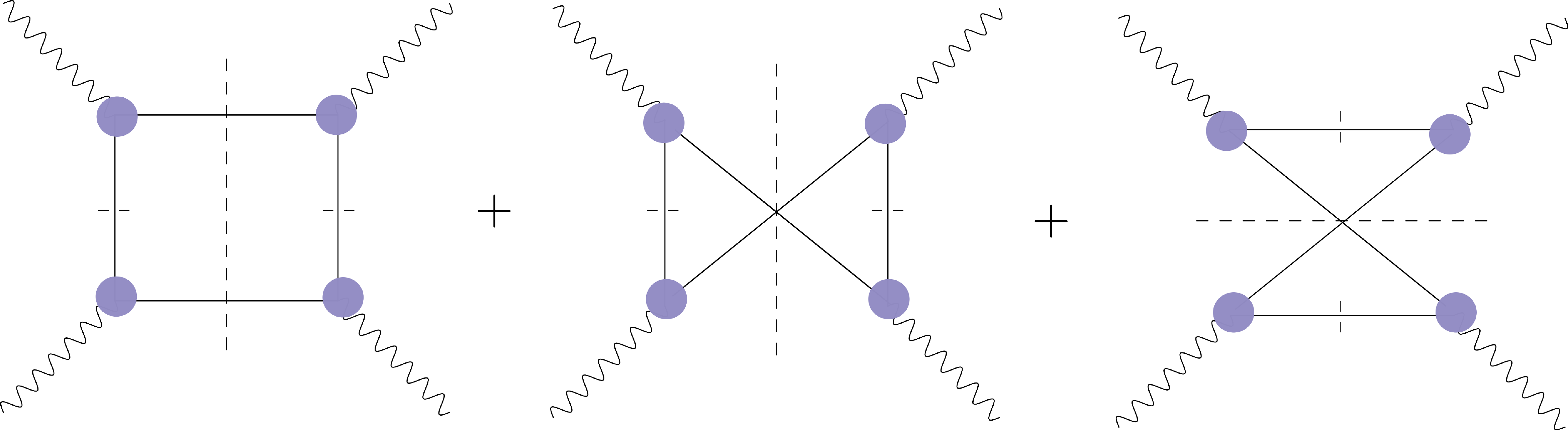}}
\caption{Leading order $\textbf{P}$-box contributions to $a_\mu^{\mathrm{HLbL}}$, where the corresponding $\textbf{P}$ meson EFFs are highlighted by the purple  filled circles.}
\label{Fig:HLBL}     
\end{figure*}

In this work we focus on a specific HLbL contribution to $a_\mu$, namely, the $\textbf{P}$-box contributions ($\textbf{P}=\pi^\pm,\,K^\pm$), depicted in Fig.~\ref{Fig:HLBL} and denoted herein as $a_{\mu}^{\textbf{P}-box}$. For pion, the dispersive evaluation of Ref.~\cite{Colangelo:2017fiz} achieved a  considerably small uncertainty of $0.2\times10^{-11}$. The Dyson-Schwinger equations (DSE) computation in Ref. \cite{Eichmann:2019bqf}  also yields a $0.02\times10^{-11}$ overall error for the $K^\pm$ contribution. This  level of accuracy serves as the reference for our corresponding computation for each case. To calculate the $\textbf{P}$-box contributions, we employ the master formula derived in~\cite{Colangelo:2017fiz}, which reads:
\begin{equation}
a_{\mu}^{\textbf{P}-box} = \frac{\alpha^3_{em}}{432 \pi^2} \int_{\Omega}  \sum_i^{12} T_i(Q_1,Q_2,\tau) \bar{\Pi}_i^{\textbf{P}-box}  (Q_1,Q_2,\tau),
\label{eq.master_formula}
\end{equation}
where $\alpha_{em}$ is the QED coupling constant and $\int_{\Omega}$ denotes the integration over the photon momenta, $Q_{1,2}$, and their relative angle $\tau$. With $Q_3^2=Q_1^2+Q_2^2+2|Q_1|  |Q_2|\tau$, the functions $\bar{\Pi}_i^{\textbf{P}-box}$ are expressed as:
\begin{eqnarray}
\label{I_feynman}
\bar{\Pi}_i^{\textbf{P}-box}(Q_1^2,Q_2^2,Q_3^2) &=& F_{\textbf{P}}(Q_1^2)F_{\textbf{P}}(Q_2^2) F_{\textbf{P}}(Q_3^2)\\
&\times& \frac{1}{16 \pi^2} \int_0^1 dx \int_0^{1-x} dy I_i(x,y)\,;\nonumber
\end{eqnarray}
the scalar functions $T_i$ and $I_i$ are provided in Appendices B and C, respectively, of  Ref.~\cite{Colangelo:2017fiz}. Thus, the only missing  ingredients are the electromagnetic form factors (EFFs), $F_{\textbf{P}}(Q^2)$, obtained from the process $\gamma^* \textbf{P} \to \textbf{P}$. 

Our approach is based upon the DSE formalism~\cite{Roberts:1994dr,Sanchis-Alepuz:2017jjd,Fischer:2018sdj}, which captures the nonperturbative character of QCD excellently well and has produced a plethora of hadron physics predictions; for instance, it unifies the description of the EFFs~\cite{Miramontes:2021xgn,Eichmann:2019bqf, Chen:2018rwz,Gao:2017mmp,Chang:2013nia} with their corresponding distribution amplitudes and distribution functions~\cite{Ding:2019qlr,Cui:2020tdf,Shi:2014uwa,Chang:2013pq}, as well as with the $\gamma^* \gamma^* \to \{\pi^0,\,\eta,\,\eta',\,\eta_c,\,\eta_b \}$ transition form factors (TFFs)~\cite{Raya:2015gva,Raya:2016yuj,Ding:2018xwy,Raya:2019dnh,Eichmann:2017wil}. This manuscript is organized as follows: Sec. II describes the computation of EFFs within the DSE formalism, dissecting all the pieces entering the corresponding electromagnetic current. The numerical results of the EFFs and contributions to $a_\mu$ are presented in Sec. III.  Section IV summarizes our results and conclusions.

\section{Electromagnetic form factors in the DSE formalism}
The interaction of a virtual photon with a pseudoscalar meson is described by a single form factor, $F_\textbf{P}(Q^2)$. The matrix element reads
\begin{equation}
\label{eq:current}
\langle\textbf{P}(p_f)|j_\mu|\textbf{P}(p_i)\rangle=2K_\mu F_\textbf{P}(Q^2)\;,
\end{equation}
where $Q=p_f -p_i$ is the photon momentum and $2K = (p_f+p_i$); the electromagnetic current is\begin{equation}
\label{eq:current2}
j_{\mu} = \bar{\Gamma}^f_\textbf{P} G_0 (\mathbf{\Gamma}_{\mu} - \mathcal{K}_{\mu}) G_0 \Gamma^i_\textbf{P} \; ,
\end{equation} 
with $\Gamma^{i,f}_\textbf{P}$ denoting the incoming and outgoing $\textbf{P}$ meson Bethe-Salpeter amplitudes (BSAs), respectively; $G_0$ represents an appropriate product of dressed quark propagators, such that 
\begin{equation}
\mathbf{\Gamma}_{\mu} = \left(S^{-1} \otimes S^{-1} \right)_{\mu} = \Gamma_{\mu}\otimes S^{-1} + S^{-1}\otimes \Gamma_{\mu}\;
\end{equation}
defines the impulse approximation (IA)~\cite{Maris:2000sk}; this will be shown explicitly later. Beyond IA effects are encoded in $\mathcal{K}_\mu$ (see appendix), which characterizes the interaction of the photon with the Bethe-Salpeter kernel describing the two-body interaction~\cite{Miramontes:2019mco,Miramontes:2021xgn}. Thus, all the parts entering Eq.~\eqref{eq:current} require the knowledge of quark propagators, BSAs, quark-photon vertex (QPV), and their corresponding interaction kernels. We shall now describe how to gather those ingredients within the DSE approach.

\subsection{Quark propagator and meson Bethe-Salpeter amplitudes}
The DSEs are the QCD equations of motion,  
encoding full dynamics of the theory,  simultaneously capturing the perturbative and nonperturbative facets of QCD~\cite{Roberts:1994dr,Fischer:2018sdj}. The DSEs form an infinite set of coupled integral equations that relate the theory's Green functions; subsequently, any tractable problem demands a systematic and rigorous truncation scheme~\cite{Qin:2020rad,Binosi:2016rxz,Eichmann:2016yit,Huber:2018ned}.

The DSE for the $f$-flavor quark propagator, also referred to as the gap equation, reads as:
\begin{eqnarray}\nonumber
    S_f^{-1}(p)&=&Z_2 [S_f^{(0)}(p)]^{-1} + \int_{q}^\Lambda [\textbf{K}^{(1)}(q,p)] S_f(q)\;,\\\label{eq:quarkPropQCD}
    [\textbf{K}^{(1)}(q,p)]&=&\frac{4}{3} Z_1 g^2 D_{\mu\nu}(p-q) [\gamma_\mu\otimes  \Gamma_\nu^{fg}(p,q)]\;,
\end{eqnarray}
where $\int_{q}^\Lambda = \int^\Lambda \frac{d^4q}{(2\pi)^4}$ stands  for  a  Poincar\'e  invariant  regularized  integration,   $\Lambda$  being the regularization scale. The  components that constitute the one-body kernel, $[\textbf{K}^{(1)}]$, carry their usual meanings (color indices have been omitted for the simplicity of notation):
\begin{itemize}
    \item $D_{\mu\nu}$ is the gluon propagator and $g$ is the coupling constant  for all the QCD interactions appearing in the Lagrangian.
    \item $\Gamma_\nu^{fg}$ represents the fully-dressed quark-gluon vertex (QGV); in general characterized by 12 Dirac structures~\cite{Albino:2021rvj,AtifSultan:2018end,  Albino:2018ncl}. 
    \item $Z_{1,2}$ are the QGV and quark wave-function renormalization constants, respectively.
\end{itemize}
Herein, $S_f^{(0)}(p)=[i \gamma \cdot p+m_f^{\text{bm}}]^{-1}$ is the bare propagator and $m^{\text{bm}}_f$ the bare fermion  mass. The fully dressed quark propagator is represented as
\begin{equation}
    \label{eq:quarkPropDef}
    S_f(p)=  Z_f(p^2) (i \gamma \cdot p + M_f(p^2))^{-1}\;,
\end{equation}
in clear analogy with its bare counterpart. Multiplicative renormalization entails that the quark mass function, $M_f(p^2)$, is independent of the renormalization point $\zeta$.

The description of mesons is obtained from the Bethe-Salpeter equation (BSE)~\cite{Eichmann:2016yit,Binosi:2016rxz,Qin:2020rad}:
\begin{equation}
\label{eq:BSEGen}
    \Gamma_{H}(p;P)=\int_{q}^\Lambda [\textbf{K}^{(2)}(q,p;P)]\chi_{H}(q;P)\;,
\end{equation}
whose ingredients are defined as follows:
\begin{itemize}
    \item As before, $\Gamma_{H}$ denotes the BSA, with $H$ labeling the type of meson. \item $\chi_{H}(q;P)=S(q_+)\Gamma_{H}(q;P)S(q_-)$ corresponds to the Bethe-Salpeter wave function (BSWF).
    \item The kinematic variables: $P$ is the total momentum of the  bound state such that $P^2=-m_H^2$ ($m_H$ the mass of the meson); $q_+=q+\eta P$ and $q_-=q-(1-\eta)P$, where $\eta\in[0,1]$ determines the relative momentum. 
\end{itemize}
The Dirac structure characterizing the BSA depends on the meson's quantum numbers. For a pseudoscalar meson $\textbf{P}$:
    \begin{eqnarray}
\label{eq:BSA}
    \Gamma_{\textbf{P}}(q;P) &=& \gamma_5[i \mathbb{E}_{\textbf{P}}(q;P)+\gamma \cdot P \mathbb{F}_{\textbf{P}}(q;P) \\
    &+& \gamma \cdot q \mathbb{G}_{\textbf{P}}(q;P) + q_\mu \sigma_{\mu \nu} P_\nu \mathbb{H}_{\textbf{P}}(q;P)]\;.\nonumber
\end{eqnarray}
The two-body interaction in Eq.~\eqref{eq:BSEGen} is represented by $[\textbf{K}^{(2)}(q,p;P)]$; it corresponds to two-particle irreducible quark/antiquark scattering kernel, which  contains all possible interactions between the quark and antiquark within the bound state~\cite{Qin:2020jig}.  Once the 1 and 2-body kernels have been specified (i.e., a truncation scheme has been defined), gap and Bethe-Salpeter equations can be solved. In fact, $[\textbf{K}^{(1)}]$ and $[\textbf{K}^{(2)}]$ are related via vector and axial-vector Ward-Green-Takahashi identities (WGTIs)~\cite{Xing:2021dwe,Qin:2014vya, Bhagwat:2007ha}, implying charge conservation and the appearance of pions and kaons (in the chiral limit) as Nambu-Goldstone bosons of dynamical chiral symmetry breaking~\cite{Bender:1996bb}. 

\subsection{Rainbow ladder truncation}
The simplest truncation that fulfils vector and axial-vector WGTIs is defined by the kernel ( $\{\,r,\,s,t,\,u\}$ are color indices):
\begin{eqnarray}
\label{eq:defRL}
    [\textbf{K}_{tu}^{rs}]^{\text{RL}}(q,p;P)=-\frac{4}{3} Z_2^2 D_{\mu\nu}^{\text{eff}}(p-q)[\gamma_\mu]_{ts} \otimes [\gamma_\nu]_{ru} \;,\hspace{0.5cm}
\end{eqnarray}
which relate the 1-body and 2-body kernels as:
\begin{equation}
\label{eq:defRL2}
    \textbf{K}^{(2)}(q,p;P)=\textbf{K}^{\text{RL}}(q,p;P)=-\textbf{K}^{(1)}(q,p;P)\;.
\end{equation}
This truncation is dubbed as the RL truncation~\cite{Bender:1996bb}.  It provides a reliable and practical approach so long as we restrain ourselves to ground-state pseudoscalar and vector mesons~\cite{Xu:2019ilh,Ding:2019qlr,Raya:2015gva,Chang:2013nia}. It is worth noticing that the gluon propagator has been demoted to an effective one, $g\,D_{\mu\nu} \to D_{\mu\nu}^{\text{eff}}$, where:
\begin{equation}
    D_{\mu\nu}^{\text{eff}}(k) =\left( \delta_{\mu \nu}-\frac{k_\mu k_\nu}{k^2} \right)\mathcal{G}(k^2)\;.
\end{equation}
Herein, $\mathcal{G}(k^2)$ is an effective coupling, typically obtained from either lattice QCD or phenomenological models~\cite{Serna:2018dwk,Chang:2021vvx,Qin:2011dd}. Throughout this work, we shall employ the well-known Qin-Chang (QC) interaction~\cite{Qin:2011dd}:
\begin{equation}\label{eq:QCmodel}
\mathcal{G}(q^2) {}=
\frac{ 8 \pi^2}{\omega^4} D
e^{-\frac{q^2}{\omega^2}}+{}\frac{2\pi\gamma_m
\big(1-e^{-q^2/\Lambda_{t}^2}\big)}{\textnormal{ln}[e^2-1+(1+q^2/\Lambda_{\text{QCD}}
^2)^2]}~.
\end{equation}
The first term above controls the strength of the effective coupling, in such a way that the QC model is defined once the mass parameter, $m_G = (w D)^{1/3}$, is fixed to produce the masses and decay constants of the ground-state pseudoscalar mesons. Typical RL parameters are $m_G \sim 0.8$ GeV and $w \sim 0.5$ GeV; herein, the later is varied within the range $w\in(0.4,\,0.6)$ to estimate model uncertainties. The second term is simply set to reproduce the 1-loop behavior of the QCD's running coupling: $\gamma_m=12/(11N_C-2N_f)=12/25$ is the anomalous dimension, with $N_f=4$ flavors and $N_c=3$ colors, and $\Lambda_{\text{QCD}}=0.234$ GeV; the parameter $\Lambda_{t}=1$ GeV is introduced for technical reasons and has no material impact on the computed observables. Table~\ref{Tab:RLparams2} collects the RL inputs and some static properties of the pion and kaon.

It is worth mentioning that the RL truncation is self-consistent with the IA,  in such a way that the EFF is obtained from:
\begin{eqnarray}
    \label{eq:defEFF}
   & 2K_\mu F_{\textbf{P}}(Q^2) = e_u [F_{\textbf{P}}^{u}(Q^2)]_\mu + e_{\bar{h}} [F_{\textbf{P}}^{\bar{h}}(Q^2)]_\mu\;,
\end{eqnarray}
where $\textbf{P}$ is a $u \bar{h}$ meson and $e_{u,\bar{h}}$ are the electric charges of the quark and antiquark, respectively. $[F_{\textbf{P}}^f(Q^2)]_\mu$ denotes the interaction of the photon with a valence constituent $f$-in-$\textbf{P}$, such that:
\begin{eqnarray}
    [F_{\textbf{P}}^f(Q^2)]_\mu =\nonumber \text{tr}_{CD} \int_q \chi_{\mu}^f(q+p_f,q+p_i)\\
    \times \Gamma_{\textbf{P}}(q_i;p_i)S(q)\Gamma_{\textbf{P}}(q_f;-p_f)\;. \label{eq:defEFF2}
\end{eqnarray}
The kinematics is defined as follows: $p_{i,f}=K\mp Q/2$ and $q_{i,f}=q+p_{i,f}/2$, such that $p_{i,f}^2=-m_{\textbf{P}}^2$; naturally, $m_{\textbf{P}}$ is the mass of the pseudoscalar meson and $Q$ the photon momentum. The trace, $\text{tr}_{CD}$, is taken over color and Dirac indices. The only remaining ingredient to compute the EFFs in the RL approximation is the QPV. This is described below.

\begin{table*}[t!]
\centering
\begin{tabular}{|l||l|l|l||l|l|l||l|l|l|}
\hline
&$m_u$ & $m_s$ & $m_G$    & $m_{\pi}$ & $f_{\pi}$ & $r_{\pi}$ & $m_k$ & $f_k$ & $r_k$  \\ \hline
RL  & 0.0052 & 0.122 & 0.80 &  0.139 (3)      &  0.131 (2)   &  0.677 (4)     &  0.493 (4)    & 0.157 (2)     &     0.597 (3)   \\ \hline
BRL & 0.0060 & 0.125& 0.84 &   0.139 (4)    &  0.131 (2)     &  0.676 (2)     &  0.493 (5)   &    0.159 (2) &      0.593 (2)      \\ \hline
\end{tabular}
	\caption{\label{Tab:RLparams2} RL and BRL parameters ($m_u=m_d$, $m_s$ and $m_G$), fixed to produce the ground-state masses and decay constants (quoted in GeV). Physical observables exhibit mild sensitivity to the variation of $w^{\text{RL}}\in(0.4,\,0.6)$ GeV and $w^{\text{BRL}}\in(0.6,\,0.8)$ GeV; this variation, however, is taken into account for the calculation of the electromagnetic form factors. Charge radii (in fm) are obtained from Eq.~\eqref{eq:ChargeRadii}.}
\end{table*}

\subsection{Quark-photon vertex}
The QPV might be obtained via the inhomogeneous BSE:
\begin{equation}
\label{eq:BSEin}
    \Gamma_{\mu}^f(p;P)=\gamma_\mu+\int_{q}^\Lambda [\textbf{K}^{(2)}(q,p;P)]\chi_\mu^f(q;P)\;,
\end{equation}
where  $\chi_\mu^f(q;P)$ is simply the unamputated vertex, 
\begin{equation}
    \chi_\mu^f(q;P)=S^f(q_+)\Gamma_{\mu}^f(q;P)S^f(q_-)\;.
\end{equation}
The choice of the 2-body kernel in Eq.~\eqref{eq:BSEin} renders the QPV self-consistent with the chosen truncation, ensuring, for example, that the Abelian anomaly related with the process $\gamma \gamma \to \pi^0$ is faithfully reproduced~\cite{Eichmann:2019tjk,Maris:2002mz}. For the purpose of clarity, we refer to this approach as the direct computation.

In the RL truncation, vector meson bound states appear as poles on the negative real axis in the $Q^2$ plane in the inhomogeneous BSE for the QPV~\cite{Maris:1999bh} and, as a consequence, in the timelike form factors. The appropriate inclusion of these poles favors obtaining the correct value for the charge radius~\cite{Maris:1999bh,Maris:2000sk}. The EFFs in the timelike region are harder to describe in the DSE-BSE approach. 
For all practical purposes,
this should not affect the way EFFs contribute to $a_\mu$, because only a relatively small spacelike region of the corresponding form factors near $Q^2=0$ actually matters for determining their contribution~\cite{Eichmann:2019bqf,Raya:2019dnh}.  We expect this small effect to be virtually remedied by adjusting the model parameters to reproduce 
the correct value of the charge radius.

Notwithstanding, it is  worth exploring and reassuring our expectations through a proper treatment of the timelike region. In order to shift the vector meson poles appearing in the QPV to the complex plane, and turn the boundstate into a resonance with a nonvanishing decay width, the interaction kernels $\textbf{K}^{(1,\,2)}$ must allow virtual decays into suitable channels~\cite{Fischer:2007ze, Fischer:2008sp}. The truncation explored in~\cite{Miramontes:2019mco} and employed in~\cite{Miramontes:2021xgn} for the calculation of the pion timelike EFF, denoted herein as {\em beyond rainbow-ladder} (BRL), takes into account resonance effects and incorporates meson cloud effects (MCEs) in the description of the pion EFF. This is sufficient to produce the correct behavior of the pion EFF in the timelike axis. We adapt  this approach to compute the $\pi-K$ EFFs and corresponding box contributions. As Eq.~\eqref{eq:current2} suggests, it is also desirable to go beyond the IA. Nevertheless, to alleviate the numerical calculations we neglect further photon couplings and consider the IA only. Some aspects of the calculation of EFFs in the BRL truncation are canvassed herein, in Appendix A, and detailed through Refs.~\cite{Miramontes:2019mco,Miramontes:2021xgn}. As clarified in Table~\ref{Tab:RLparams2}, the QC model favors $m_G=0.87$ GeV and $\omega \sim 0.7$ GeV.

Due to technical reasons, when employing the QPV obtained from Eq.~\eqref{eq:BSEin}, the calculation of EFF is limited to a certain domain of spacelike momenta. For instance, the pion elastic and $\gamma^* \gamma \to \pi^0$ transition form factors can only be obtained up to $Q^2\sim4$ GeV$^2$~\cite{Maris:2000sk,Maris:2002mz}, without appealing to sophisticated mathematical techniques for extrapolation~\cite{Eichmann:2017wil}. While not the entire spacelike domain is crucial to $a_\mu$, it is reassuring to access it in its entirety~\footnote{The large-$Q^2$ behavior of the $\gamma^* \gamma^*$ TFFs is quite useful to parametrize the numerical solutions~\cite{Knecht:2001qf,Raya:2019dnh}.}. For this  reason and in
direct connection with our previous work on the HLbL contributions of neutral pseudoscalars~\cite{Raya:2019dnh}, we also present an alternative technique, based upon perturbation theory integral representations (PTIRs), to evaluate the form factors at arbitrarily large momenta.

\subsection{The PTIR approach}
A practical  PTIR approach for the quark propagators and BSAs was put forward in~\cite{Chang:2013pq,Chang:2013nia}, to calculate the pion distribution amplitude and spacelike EFF.  It was subsequently implemented to the case of $\gamma^* \gamma^*$ TFFs~\cite{Raya:2015gva,Raya:2016yuj,Ding:2018xwy,Raya:2019dnh}. The general idea, which applies to all pseudoscalars, is to describe the quark propagators in terms of $j_m=2$ complex conjugate poles (CCPs), and express the BSAs, $\mathcal{A}_j$, as follows:
	\begin{subequations}\label{BSAPTIR}
		\begin{eqnarray}
		\mathcal{A}_j(k;P)= \sum_{i=1}^{i_n} \int_{-1}^1 dw \rho_i^j(w) \frac{c_i^j \, (\Lambda^2_{i,j})^{\beta_i^j}}{(k^2+w k \cdot P + \Lambda^2_{i,j})^{\alpha_i^j}}\;.\hspace{0.7cm}\nonumber\\
		\end{eqnarray}
	\end{subequations}
The interpolation parameters involved, i.e., $\{z_j, m_j \}$, $\{\alpha_i^j, \beta_i^j, \Lambda_{i,j}, c_i^j, i_n = 3 \}$ (for quark propagators and  BSAs, respectively), as well as the spectral weights, $\rho_i^j(w)$, are determined through fitting of the numerical results of the corresponding DSE-BSEs. The  carefully constructed sets of RL truncation parameters are found in Refs.~\cite{Raya:2015gva,Shi:2014uwa}.

Constructing a PTIR for the QPV in Eq.~\eqref{eq:BSEin} turns out to be difficult and unpractical~\cite{Xu:2019ilh}. Thus, appealing to gauge covariance properties~\cite{Delbourgo:1977jc}, the following Ansatz has been proposed and systematically tested~\cite{Raya:2015gva,Raya:2016yuj,Ding:2018xwy,Raya:2019dnh}:
	\begin{eqnarray}
	\nonumber
	\chi_\mu^f(k_o,k_i)&=& \gamma_\mu \Delta_{k^2 \sigma_V}^f \\
	\nonumber
	&+& [\mathbf{s}_f \gamma\cdot k_o \gamma_\mu \gamma \cdot k_i + \bar{\mathbf{s}}_f\gamma\cdot k_i \gamma_\mu \gamma \cdot k_o]\Delta_{\sigma_V}^f\nonumber\\
	\nonumber
	&+&[\mathbf{s}_f(\gamma\cdot k_o \gamma_\mu + \gamma_\mu \gamma \cdot k_i)\\
	&+&\bar{\mathbf{s}}(\gamma\cdot k_i \gamma_\mu + \gamma_\mu \gamma \cdot k_o)]i\Delta_{\sigma_S}^f\;,
	\label{eq:vertex}
	\end{eqnarray}
	where $\Delta_\phi^f=[\phi^f(k_o^2)-\phi^f(k_i^2)]/(k_o^2-k_i^2)$ and $\bar{\mathbf{s}}_f=1-\mathbf{s}_f$.  According to ~\cite{Raya:2019dnh}, the transverse pieces are weighted by
	\begin{equation}
	\label{eq:TTs}
	\mathbf{s}_f = \mathbf{s}_f\; \textrm{exp}\left[-\left(\sqrt{Q_1^2/4 + m_\textbf{P}^2}-m_\textbf{P}\right)/M^E_f\right]\;,
	\end{equation}
such that the strength parameter $\textbf{s}_f$ is tuned to reproduce the $\pi^0$ Abelian anomaly and $\gamma \gamma \to \{\eta,\,\eta',\,\eta_c\}$ empirical decay widths. Nonetheless, as confirmed by our numerical evaluations, such weighting for the case of the EFFs is irrelevant and one can simply set $\textbf{s}_f$ to zero. The reasons can be easily understood: first, the terms which dominate at low-$Q^2$ in Eq.~\eqref{eq:defEFF2}, those involving the product of leading BSAs ($\mathbb{E}_\textbf{P} \times \mathbb{E}_\textbf{P}$), are not affected at all by the choice of $\textbf{s}_f$ since the corresponding trace is exactly zero; then, with $F_\textbf{P}(Q^2=0)=1$ entirely fixed by charge conservation, being exponentially suppressed, the $\textbf{s}_f$-weighted subleading terms could only provide a minor contribution in neighborhood of $Q^2\sim 0$.

Defined as in Eq.~\eqref{eq:vertex}, the QPV is fully written in terms of the quark propagator dressing functions~\footnote{The $f$-quark propagator expressed $S_f(p) = -  i \gamma \cdot p\;\sigma_v^f(p^{2}) + \sigma_s^f(p^{2})$, with $\sigma^f_{s,v}(p^2)$ being algebraically related to $M_f(p^2)$ and $Z_f(p^2)$ in Eq.~\eqref{eq:quarkPropDef}.}. With all the ingredients in Eqs.~(\ref{eq:defEFF}, \ref{eq:defEFF2}) expressed in a PTIR, the evaluation of the 4-momentum integral follows after a series of standard algebraic steps (numerical integration is only carried out for the Feynman parameters and spectral weights). Hence, the form factors can be calculated at arbitrarily large spacelike momenta.

\section{Numerical Results}
\subsection{Electromagnetic form factors}
\begin{figure}[htb]
 \centering
 \includegraphics[width=\linewidth]{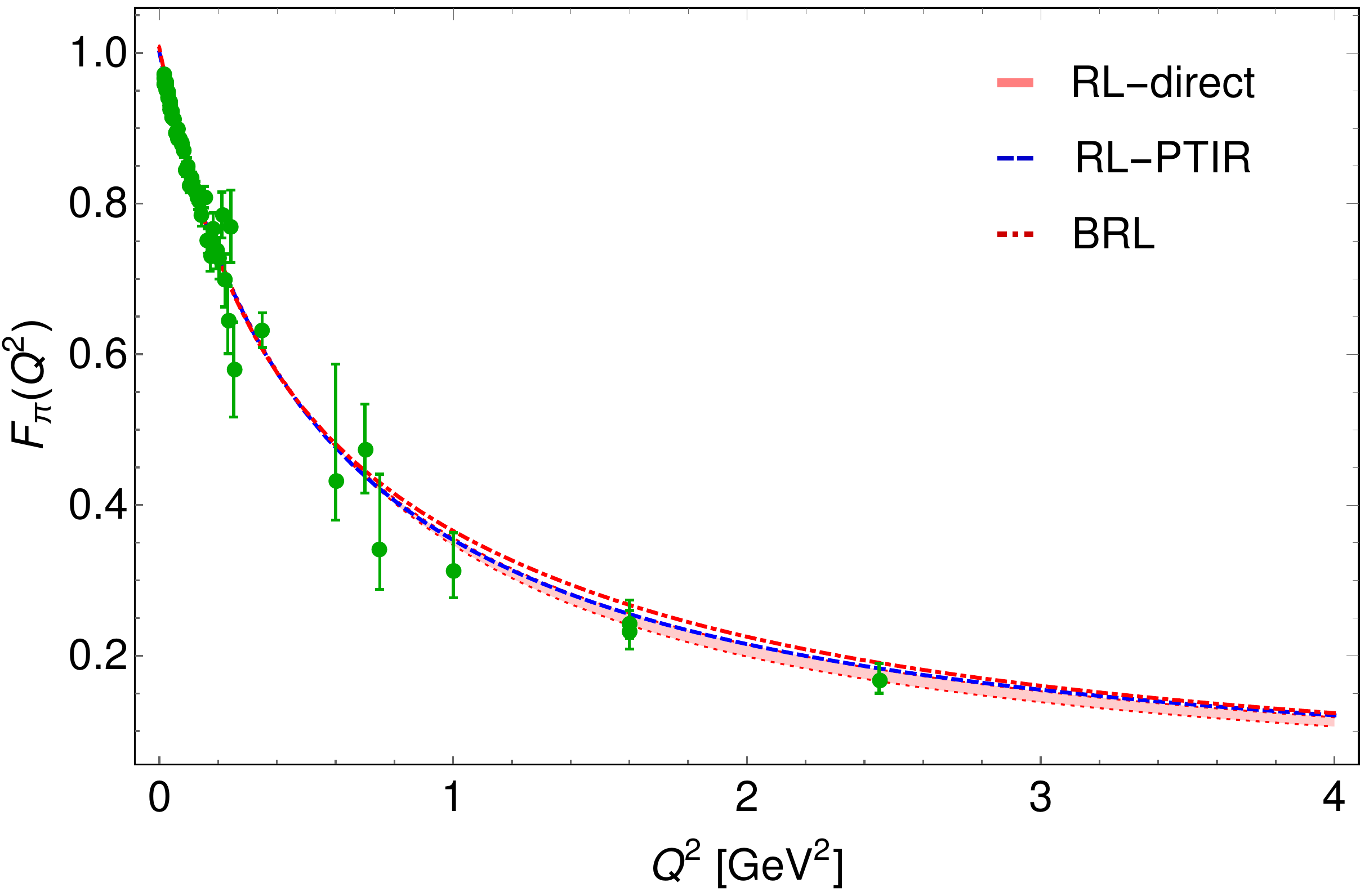}\\
 \includegraphics[width=\linewidth]{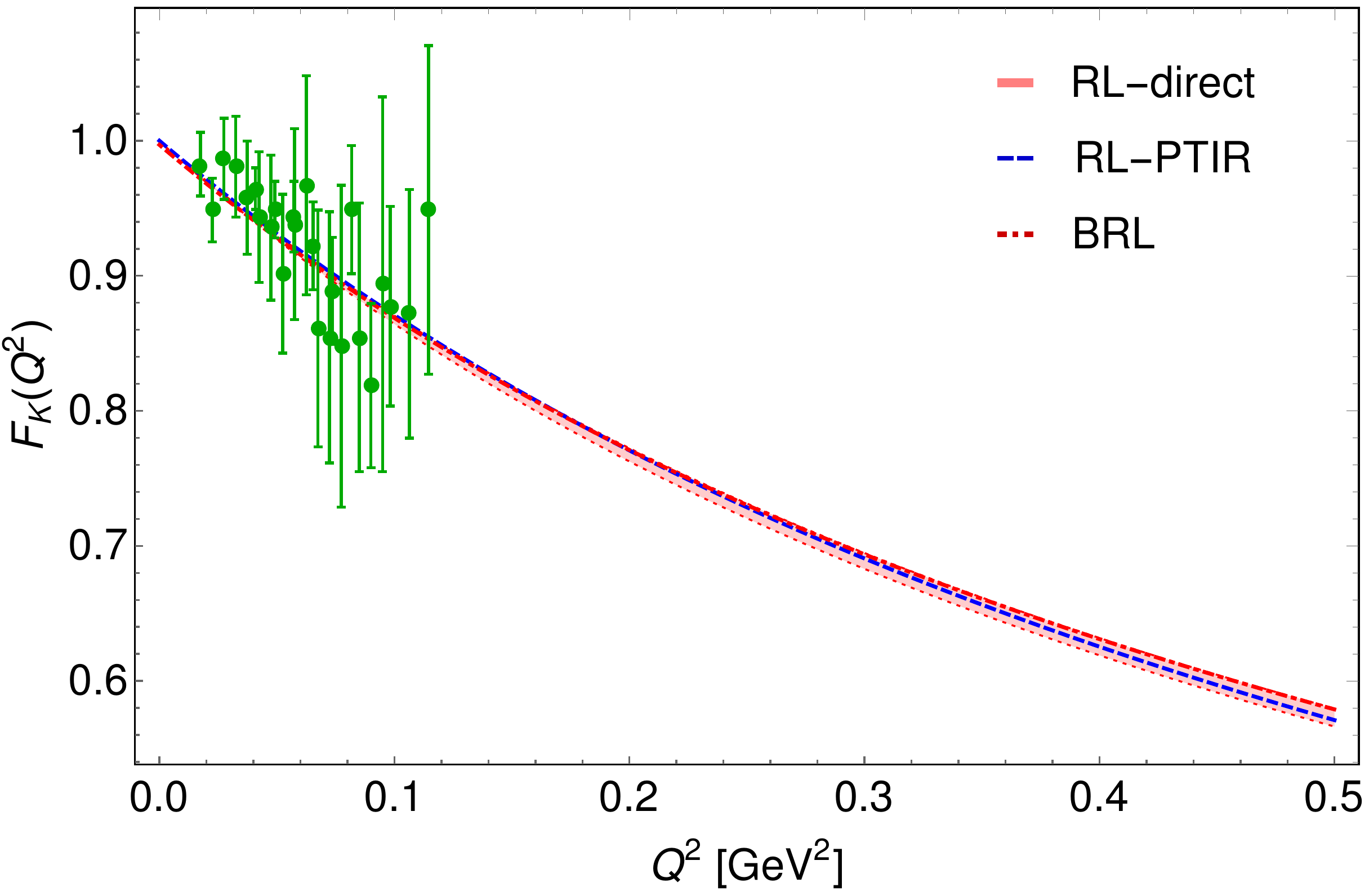}\\
 \caption{ $\pi^+$ and $K^+$ EFFs. The narrow band in the RL-direct result accounts for the variation of the QC model parameters, as described in text; those corresponding to the PTIR and BRL results are not shown, since there is a considerable overlap. The charge radii, Table~\ref{Tab:RLparams2}, are practically insensitive to the model inputs and truncation. Experimental data is taken from Refs.~\cite{JeffersonLab:2008gyl,Dally:1980dj,Amendolia:1986ui,NA7:1986vav}.}
 \label{fig:EFFs}
\end{figure}

The $\pi-K$ EFFs are presented in Fig.~\ref{fig:EFFs}. We compare the RL results which follow from the direct computation and PTIR approach; the compatibility between both calculations is evident. In the domain of interest, the BRL truncation yields similar outcomes. Furthermore, our obtained EFFs are in clear agreement with the DSE results reported in Ref.~\cite{Eichmann:2019bqf}.  The charge radii are obtained from the derivative of the form factor:
\begin{equation}
    \label{eq:ChargeRadii}
    r_\textbf{P}^2 = -6 \frac{dF_\textbf{P}(Q^2)}{dQ^2}\big|_{Q^2=0}\;.
\end{equation}
Both RL and BRL direct computations yield similar values: $r_\pi = 0.677(4)$ fm and $r_K=0.597(3)$ fm (RL), and $r_\pi = 0.676(2)$ fm and $r_K=0.593(2)$ fm (BRL); the error accounts for the variation of $\omega$ in the QC model, as explained in Table~\ref{Tab:RLparams2}. The RL-PTIR case also falls within these values:  $r_\pi = 0.676(5)$ fm and $r_K=0.596(5)$. 


\subsection{Pion and kaon box contributions}
The integrations in Eqs.~(\ref{eq.master_formula}-\ref{I_feynman}) have been carried out employing the CUBA library~\cite{Hahn:2004fe}, benchmarked with the vector meson dominance (VMD) ans\"atze of the form factors:
\begin{eqnarray}\label{eq:VMD}
F_{\pi^+}^{\text{VMD}}(Q^2) &=& \frac{m_\rho^2}{m_{\rho}^2 + Q^2}\,,\\ \nonumber
F_{K^+}^{\text{VMD}}(Q^2) &=& 1 - \frac{Q^2}{2} \Bigg[\frac{1}{m_{\rho}^2 + Q^2} + \frac{1}{3} \left(\frac{1}{m_{\omega}^2 +Q^2} \right)\\
&+&\frac{2}{3} \left(\frac{1}{m_{\phi}^2 + Q^2}\right) \Bigg]\,,
\end{eqnarray}
which yield the results
\begin{eqnarray} \nonumber
a_\mu^{\pi^{\pm}-\text{box}} &=& -
16.4
\times 10^{-11} \nonumber \;\;\text{[RL-PTIR]}\;, \\
a_\mu^{K^{\pm}-\text{box}} &=& -
0.5
\times 10^{-11} \;\;\text{[BRL]}\;, \label{eq:boxpikVMD}
\end{eqnarray}
where we have employed $m_{\pi} = 0.13957$ GeV, $m_K = 0.49367$ GeV, $m_{\rho} = 0.7752$ GeV, $m_\omega = 0.7827$ GeV, $m_\phi=1.0195$ GeV, $m_{\mu} = 0.10565$ GeV and $\alpha_{em} = 1/137.03599$~\cite{ParticleDataGroup:2020ssz}. The estimates in Eq.~\eqref{eq:boxpikVMD} match those quoted in~\cite{Eichmann:2019bqf}, and the integration errors have been omitted, since those are two orders of magnitude smaller.

With the EFFs obtained in the RL (direct and PTIR) and BRL truncations, the numerical estimates for the $\pi^\pm-$box contributions are:
\begin{eqnarray} \nonumber
a_\mu^{\pi^{\pm}-\text{box}} &=& -(15.4\pm 0.3) \times 10^{-11} \;\;\text{[RL-direct]}\;, \\
a_\mu^{\pi^{\pm}-\text{box}} &=& -(15.6\pm 0.3) \times 10^{-11} \nonumber \;\;\text{[RL-PTIR]}\;, \\
a_\mu^{\pi^{\pm}-\text{box}} &=& -(15.7\pm 0.2) \times 10^{-11} \;\;\text{[BRL]}\;. \label{eq:boxpi}
\end{eqnarray}
Analogous results for the $K^\pm$ case yield:
\begin{eqnarray}
a_\mu^{K^{\pm}-\text{box}} &=& -(0.47\pm 0.03) \times 10^{-11} \nonumber \;\;\text{[RL-direct]}\;, \\
a_\mu^{K^{\pm}-\text{box}} &=& -(0.48\pm 0.03) \times 10^{-11} \nonumber \;\;\text{[RL-PTIR]}\;, \\
a_\mu^{K^{\pm}-\text{box}} &=& -(0.48 \pm 0.02) \times 10^{-11} \;\;\text{[BRL]}\;.\label{eq:boxkaon}
\end{eqnarray}

From Fig.~\ref{fig:EFFs} and the above estimates, it is clear that the direct and PTIR approach are  virtually indistinguishable; the BRL truncation also yields similar outcomes.  Therefore, one can combine the estimates in Eqs.~\eqref{eq:boxpi}-\eqref{eq:boxkaon} to produce:
\begin{eqnarray}
    a_\mu^{\pi^\pm-\text{box}}&=&-(15.6\pm 0.2)\times 10^{-11}\;,\\
    a_\mu^{K^\pm-\text{box}}&=&-(0.48\pm 0.02)\times 10^{-11}\;,
\end{eqnarray}
where the  weighted errors have been added
.

Our result for the $\pi^\pm$-box contribution agrees remarkably with the dispersive one, $-15.9(2)\times10^{-11}$~\cite{Colangelo:2017fiz} and an earlier DSE evaluation, $-15.7(2)(3)\times10^{-11}$~\cite{Eichmann:2019bqf} (see also Ref.~\cite{Eichmann:2019tjk}). In the case of the $K^\pm$-box contribution, we agree again with the previous DSE computation~\cite{Eichmann:2019bqf}, $-0.48(2)(4)\times10^{-11}$, which yields  $-0.46(2)\times10^{-11}$ once the integration error is improved~\cite{Aoyama:2020ynm}.
We note that the $K^0$-box contribution is very much suppressed, as can be seen from its VMD description in the ideal $\omega$-$\phi$ mixing case
\begin{eqnarray}\nonumber
  F_{K^0}^{\text{VMD}}(Q^2) &=&  \frac{Q^2}{2} \Bigg[-\frac{1}{m_{\rho}^2 + Q^2} + \frac{1}{3} \left(\frac{2}{m_{\omega}^2 +Q^2} \right) \\  &+&\frac{2}{3} \left(\frac{1}{m_{\phi}^2 + Q^2}\right) \Bigg]  ,
\end{eqnarray}
which yields the negligible result $\sim1\times10^{-15}$~\cite{Aoyama:2020ynm}. Consequently, we do not evaluate this contribution in our framework.

\section{Conclusions and scope}
We describe the computation of the EFFs $\gamma^* \textbf{P} \to \textbf{P}$ within the DSE approach to QCD,
leading to the evaluation of their contributions to $a_\mu$. The EFFs were obtained, firstly, in the RL truncation. Direct computations and the PTIR approach were shown to be fully compatible, while also being in agreement with the DSE results from Refs.~\cite{Eichmann:2019tjk,Eichmann:2019bqf}. Our previous calculation of the ground-state pseudoscalar pole contributions reinforces this finding~\cite{Raya:2019dnh}. It was confirmed that the BRL truncation, which incorporates meson cloud effects~\footnote{For spacelike EFFs, meson cloud effects take place in the neighborhood of $Q^2\approx 0$~\cite{Williams:1993ux,Ramalho:2016zgc,Granados:2017cib}, such that, for increasing $Q^2$, BRL $\to$ RL.}, produces similar EFFs in the relevant domain for $a_\mu$; the value of the latter being barely affected by the new effects in the truncation.

In this way, we have highlighted how the DSE formalism is a robust approach for calculations of hadronic observables, including quantities of interest for the muon $g-2$. We hope to continue developing calculations related to the subject; for instance, the importance of axial mesons has been discussed in~\cite{Roig:2019reh}, and the contribution coming from excited states might be relevant as well.

\begin{acknowledgments}
The authors acknowledge support from CONACYT and Cátedras Marcos Moshinsky (Fundación Marcos Moshinsky). This research was also supported by CONACYT grant `Paradigmas y Controversias de la Ciencia 2022' (Project No. 319395) and by Coordinación de la Investigación Científica (CIC) of the University of Michoacán, México, through Grant No. 4.10.
\end{acknowledgments}

\appendix

\section{Beyond Rainbow-ladder truncation}
\label{appendix:BRL}
In the RL truncation, the meson bound states appear as poles on the negative real axis on the $Q^2$ plane in the inhomogeneous BSE for the quark-photon vertex and, as a consequence, in the calculation of form factors in the timelike regime. 
In order to move the pole from the real axis to the complex plane and turn the bound state into a resonance state with a nonvanishing decay width, the interaction kernel $K$ must allow virtual decays into suitable channels. In Refs.~\cite{Fischer:2007ze, Fischer:2008sp}, pion cloud effects were investigated by the inclusion of pionic degrees of freedom in the quark propagator DSE and in the BSE interaction kernel. In such BRL truncation the quark propagator is modified by ($k=p-q$ and $\bar{k}=(p+q)/2$): 
\begin{eqnarray}\nonumber
S_f^{-1}(p) = S_f^{-1}(p)^{\text{RL}}\\
\label{eq:quarkDSE_tchannel}
-
\frac{3}{2}Z_2  \int_q^\Lambda \Bigg[\gamma_5 S(q) \Gamma_{\textbf{P}}\left(\bar{k}, -k\right)   + \gamma_5S(q)\Gamma_{\textbf{P}}\left(\bar{k}, k\right)\Bigg] \frac{D_{\textbf{P}}(k)}{2}~,\nonumber\\
\end{eqnarray}
with $S^{-1}(p)^{\text{RL}}$ being the right-hand-side of Eq.~\eqref{eq:quarkPropQCD} in the RL truncation, Eqs.~(\ref{eq:defRL}, \ref{eq:defRL2}), and $D_\textbf{P}(k)=(k^2+m_\textbf{P}^2)^{-1}$. The quark propagator in Eq.~\eqref{eq:quarkDSE_tchannel} preserves the axial-vector WGTI identity in combination with the following interaction kernel for the $t-$ channel pseudoscalar exchange~\cite{Miramontes:2019mco,Miramontes:2021xgn}:
\begin{eqnarray} \nonumber
\textbf{K}^{tu}_{sr}(q,p;P)&=& -\frac{3}{16} \Bigg( 
 [\Gamma_{\textbf{P}}^j]_{ru} \left(\bar{k}-P/2; k \right) [Z_2 \tau^j \gamma^5]_{ts}  \\ \nonumber
   &+&[\Gamma_{\textbf{P}}^j]_{ru} \left(\bar{k}-P/2; -k \right) [Z_2 \tau^j \gamma^5]_{ts} \\ \nonumber
   &+& [\Gamma_{\textbf{P}}^j]_{ts} \left(\bar{k}-P/2; k \right) [Z_2 \tau^j \gamma^5]_{ru} \\ \nonumber
&+& [\Gamma_{\textbf{P}}^j]_{ts} \left(\bar{k}-P/2; -k \right) [Z_2 \tau^j \gamma^5]_{ru} \Bigg)\;  D_{\textbf{P}}(k)\;.
 \label{eqn:pionkernel_1}
\end{eqnarray}
Analogous expressions for the $s$ and $u$ channels might be found in~\cite{Miramontes:2019mco}; beyond IA corrections to Eq.~\eqref{eq:current2}, $\mathcal{K}_\mu$ are given in~\cite{Miramontes:2021xgn}.

\bibliographystyle{unsrt}
\bibliography{main}

\end{document}